\documentclass[conference]{IEEEtran}
\IEEEoverridecommandlockouts
\usepackage{cite}
\usepackage{amsmath,amssymb,amsfonts}
\usepackage{graphicx}
\usepackage{textcomp}
\usepackage{xcolor}
\usepackage{multirow} 
\usepackage{subcaption}

\usepackage[utf8]{inputenc}
\usepackage{hyperref} 

\usepackage[ruled,vlined]{algorithm2e}

\usepackage{amsmath}
\usepackage{amsfonts}
\usepackage{tikz}

\newcommand{\smallhalfleft}{%
\begin{tikzpicture}[baseline=-0.3ex]
    \draw[thick] (0,0) circle (0.6ex);
    \fill (0,0.6ex) arc (90:270:0.6ex) -- cycle;
\end{tikzpicture}%
}

\newcommand{\smallfullsolid}{%
\begin{tikzpicture}[baseline=-0.3ex]
    \draw[thick] (0,0) circle (0.6ex);
    \fill[black] (0,0) circle (0.6ex); 
\end{tikzpicture}%
}

\newcommand{\smallunfilled}{%
\begin{tikzpicture}[baseline=-0.3ex]
    \draw[thick] (0,0) circle (0.6ex); 
\end{tikzpicture}%
}

\usepackage{multirow} 

\newcommand{\ignore}[1]{}

\newcommand{\linebreakand}{%
  \begin{@IEEEauthorhalign}
  \hfill\mbox{}\par
  \mbox{}\hfill\end{@IEEEauthorhalign}
}

\def\BibTeX{{\rm B\kern-.05em{\sc i\kern-.025em b}\kern-.08em
    T\kern-.1667em\lower.7ex\hbox{E}\kern-.125emX}}
\begin{document}


\title{ExplainGuard: A Zero Trust Framework for Post-Hoc Explanation Integrity Guarantees in Blackbox XAI Models}

\author{\IEEEauthorblockN{Maraz Mia}
\IEEEauthorblockA{\textit{Department of Computer Science} \\
\textit{Tennessee Tech University}\\
Cookeville, TN, USA\\
mmia43@tntech.edu}
\and
\IEEEauthorblockN{Shovan Roy}
\IEEEauthorblockA{\textit{Department of Computer Science} \\
\textit{Tennessee Tech University}\\
Cookeville, TN, USA\\
sroy42@tntech.edu}

\linebreakand

\IEEEauthorblockN{Maanak Gupta}
\IEEEauthorblockA{\textit{Department of Computer Science} \\
\textit{Tennessee Tech University}\\
Cookeville, TN, USA\\
mgupta@tntech.edu}
\and
\IEEEauthorblockN{Mir Mehedi A. Pritom}
\IEEEauthorblockA{\textit{Department of Computer Science} \\
\textit{Tennessee Tech University}\\
Cookeville, TN, USA\\
mpritom@tntech.edu}
%
}

\maketitle

\begin{abstract}
As machine learning (ML) models are increasingly deployed in high-stakes environments, explainable AI (XAI) methods like SHAP and LIME have become essential for regulatory compliance and trust. However, the current auditing paradigm relies on an implicit ``chain of trust'' where third-party auditors are assumed to be trusted. Recent research demonstrates that this assumption is flawed and adversarial auditors can manipulate XAI explanations through manipulation attacks such as {\em output shuffling} or {\em scaffolding out-of-distribution (OOD)} to conceal model biases while maintaining high prediction accuracy aiming for {\em fairwashed} explanation. In this paper, we introduce a novel defense framework, {\em ExplainGuard}, that leverages a zero trust architecture (ZTA) design to be incorporated within the XAI explanation supply chain and ensures the integrity of generated explanation. This framework would help us to replace the ambiguous default assumption of ``auditor is trustworthy,'' with a continuous ``verify-then-trust'' approach. Our design architecture establishes a Policy Decision Point (PDP) that enforces three distinct pillars of verification before any explanation is released to the user: (1) {\em asset integrity} via behavioral fingerprint to detect model substitution, (2) {\em semantic validity} using axiomatic consistency checks to reject mathematically impossible explanations, and (3) {\em feature faithfulness verification} utilizing a ranking stability approach with minimal computational overhead. Finally, we evaluate how {\em ExplainGuard} can effectively neutralize state-of-the-art explanation manipulation attacks while transforming the auditing process into a verifiable operation.

\end{abstract}

\begin{IEEEkeywords}

Secure XAI system, Zero Trust in XAI, Explainable AI, XAI security, explanation manipulation, Post-hoc Explanation Guarantees 
\end{IEEEkeywords}

\section{Introduction and Motivation}
\label{sec:intro}
The rapid adoption of ``black-box'' machine learning or deep learning models in critical sectors such as cybersecurity, finance, or healthcare has created an urgent need for explainability and transparency \cite{hosain2024explainable}. To aid that need, explainable artificial intelligence (XAI) arose from a fundamental scientific necessity with an objective of interpreting the decision outcome process for any AI-driven system outcomes \cite{singh2024towards}. Practical XAI methodologies are wide-ranging, serving as proxies to illuminate model behavior, including feature perturbation methods, notably SHAP \cite{shap_original_lundberg2017unified} and LIME \cite{lime_original_ribeiro2016should}; gradient calculation methods, such as Integrated Gradients (IG) \cite{IG_original_sundararajan2017axiomatic} and Grad-CAM \cite{selvaraju2017grad_cam_original}; simple surrogate models (e.g., Decision Tree surrogates); decomposition-based techniques, including Layer-wise Relevance Propagation (LRP) \cite{binder2016layer_LRP_original} and DeepLIFT \cite{shrikumar2017learning_deep_lift_original}; and counterfactual explanations \cite{verma2024counterfactual}. 
These explanations serve as the primary interface for stakeholders to diagnose errors, audit for fairness, and ensure regulatory compliance of any model. However, while the organizations rigorously verify the identity of users and the accuracy of models, they often have this false assumption of trust within the {\em explanation generation process} aka {\em explainability supply-chain}.


In XAI systems, explanation mechanisms are categorized as either model-agnostic or model-specific. Model-agnostic mechanisms are capable of interpreting any black-box system, while model-specific mechanisms are designed for specific model architectures. A major XAI vulnerability stems from adversarial attacks, which are engineered to deceive post-hoc explanation techniques to obscure underlying model biases, thereby obstructing an accurate assessment of the model's true readiness \cite{pachl2025view}. This threat is amplified by user behavior, as evidence suggests that users place higher reliance on the predictive performance (i.e., accuracy) of AI models than on the descriptive content provided by the accompanying explanations \cite{accuracy_and_xai_papenmier_2022}. This insight is critical because it indicates that even if the underlying explanation system is compromised, accurate predictions from the model may still lead users to trust both the output and the manipulated explanation. Since any post-hoc explanation is an approximation of the AI model's internal workings, an adversarial attack on the explanation only exacerbates this inherent approximation, and furthermore, stakeholders, who are often non-technical personnel, are predominantly focused on the model's global behavior, making them highly susceptible to such deceptive attacks.

The blind trust in {\em explainability supply-chain} presents a critical security vulnerability for various XAI methods. Recent literature has exposed that post-hoc explanation methods are highly susceptible to adversarial manipulation \cite{slack2020fooling, outputyuan2024fooling,mia2025explainable}. Thus, we propose to add an independent third-party auditor in the XAI pipeline to always verify the generated explanation. However, malicious actors acting as third-party auditors can orchestrate sophisticated attacks to deceive stakeholders. For instance, {\em fairwashing} attacks \cite{noppel2024sok, mia2025explainable} on explanation allows an adversary to manipulate feature importance scores to hide discriminatory biases (e.g., gender or race) without altering the model's predictions \cite{aivodji2019fairwashing}. Moreover, there exist no formal design architecture that treats the third-party {\em auditor} as an untrusted entity and verifies the XAI generated explanation before it reaches the user (i.e., decision-maker). This absence of verification mechanisms transforms external audits into a security liability rather than a compliance asset.

To address the above issues, in this paper we propose {\em ExplainGuard}, a novel defense framework that leverages a zero trust architecture (ZTA) within the XAI pipeline to ensure explanation integrity. To illustrate, ZTA is a cybersecurity paradigm focused on resource protection where trust is never granted implicitly \cite{stafford2020zero}. ZTA moves defenses from static perimeters to dynamic policy enforcement. Key tenets include: (1) All resource authentication is dynamic; (2) Access to resources is granted on a per-session basis; and (3) Access is determined by dynamic policy. Current security frameworks, including the NIST ZTA \cite{stafford2020zero}, focus primarily on {\em micro-segmentation}, {\em identity governance}, and {\em resource health}. While ZTA has been applied to MLOps pipelines to secure model weights \cite{zerotrustinmlops}, to the best of our knowledge, it has not been applied to the {\em output artifacts} (e.g., explanation) of XAI. 
In summary, this paper makes the following major contributions: 
\begin{itemize}
    \item Provide a novel zero trust design architecture for a verified auditing of the XAI generated explanation, replacing the traditional ``assume-trusted'' auditing model with a ``verification-based'' model.   
    
    \item Introduce a behavioral fingerprinting approach to detect manipulated explanation generated by an adversary.
    
    \item Provide a novel method to detect model substitution or adversarial wrapper model attacks in XAI setup.
\end{itemize} 
\noindent \textbf{Paper Organization.} Section \ref{sec:related-works} presents the mathematical formalization. Section \ref{sec:securexai_zta} describes the design architecture for our proposed {\em ExplainGuard} framework. Section \ref{sec:logic_proof} presents the logical proof of concept for the proposed framework. Section \ref{sec:experiments} presents experimental case study and evaluation of the framework. Section \ref{sec:limitations} discusses the limitations and future directions, while section \ref{sec:conclusion} concludes the paper.

\section{Problem Formalization and Background}
\label{sec:related-works}
\subsection{Mathematical Formulation}
Let $\mathcal{X} \to \mathbb{R}^d$ be the feature input values, $F$ being the feature column, $\mathcal{Y} \to \mathbb{R}$ be the output label space, and $F_S \subset F$ be the subspace of sensitive features. Let $x \in \mathcal{X}$ be an input instance and $f(x) \in \mathcal{Y}$ be the prediction of the machine learning model.
Let $g(x, f)$ be the explanation generated by an XAI method for the prediction outcome $f(x)$.
We define the {Adversarial Wrapper Model $f'$} when the attacker retains a closer prediction accuracy by using a wrapped or approximated version of the unfair model $f$, but for some instances, unwanted explanations are generated. To illustrate, 
$$ f \to f' \implies \left\{ \begin{array}{l} \exists x \in \mathcal{X} \quad (g(x, f) \neq g(x, f')) \\ \forall x \in \mathcal{X} \quad f(x) \approx f'(x) \end{array} \right. $$
Both $f$ and $f'$ are kept as blacbox models so, for an outside perspective, it becomes harder to identify the authenticity of the model $f$ that it was actually used to generate the explanation $g$ because of the the same predictive accuracy.



\subsection{Attacks on XAI System}
The attacks on XAI systems can be classified as (1) \textit{I}-attack or prediction preserving  \textit{(PP)} attack, where explanation is manipulated but original prediction is retained, (2) \textit{CI} or dual attack (\textit{D})-attack, where both prediction and explanation get altered \cite{kuppa2020blackbox, noppel2024sok} and (3) explanation preserving \textit{(EP)} attack where the explanation got unaltered. In the high stake scenarios, the demanding party only requires the global explanation of the model that reveals the overall feature importance for some data. A malicious auditor, whose task is to generate these explanation, can manipulate this result without affecting the predictive power of the AI model. 

We find literature where the malicious auditor deceives the target XAI method by applying wrapper or scaffold model on the original black-box model. Among such, the output shuffling attack \cite{yuan2023a, outputyuan2024fooling} utilizes an adversarial scoring function, $a(\mathcal{X})$, which is built upon a base function $f(\mathcal{X})$ that considers only non-protected features $\mathcal{X} \setminus x_\rho$ where $ x_\rho \in F_S$. This adversarial function employs a swapping function, $h_{swap}(f(\mathcal{X} \setminus x_\rho), x_\rho)$, that strategically shuffles the scores of adjacent candidates based on their protected feature $x_\rho$. This process subtly introduces bias without direct access to the input data distribution, aiming to deceive explainers like SHAP and LIME into perceiving fairwashing. Scaffolding OOD Attack \cite{slack2020fooling} conceals classifier biases from post-hoc explanation techniques like SHAP and LIME. It exploits the fact that these explainers often generate out-of-distribution (OOD) data through input perturbations. The adversary constructs an OOD detector and an unbiased classifier ($\psi$) using non-sensitive features. The core is an adversarial model ($e$) that behaves like the original biased classifier ($f$) on in-distribution data, but switches to the unbiased classifier ($\psi$) when it detects OOD probes from explainers and fools explainers into highlighting innocuous features instead of the true bias, thus fairwashing the explainer.

\section{Zero-Trust Architecture for Designing ExplainGuard Framework}
\label{sec:securexai_zta}
\subsection{Design Overview}
\begin{figure}[!htbp]
\centerline{\includegraphics[width=0.94\columnwidth]{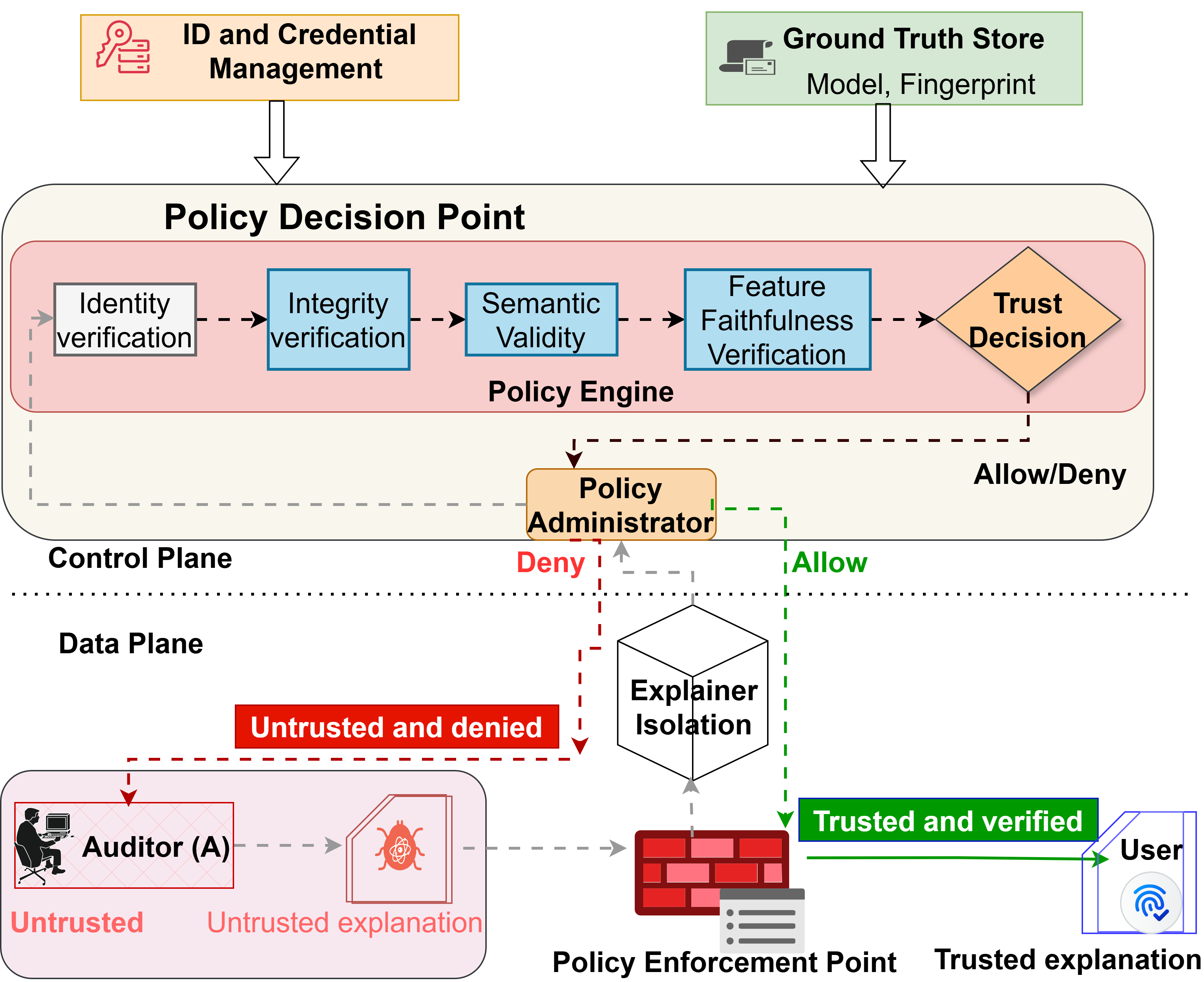}} 
\caption{Proposed zero trust architecture-based {\em ExplainGuard} framework overview}
\label{fig:architecture}
\end{figure}

We propose {\em ExplainGuard}, a Zero Trust Architecture (ZTA) designed to secure the XAI supply chain depicted in figure \ref{fig:architecture}. Unlike traditional auditing frameworks that rely on the perimeter-based trust of an authenticated auditor, {\em ExplainGuard} enforces a data-centric security model. It treats the generated explanation artifact as an untrusted object that must be continuously verified for its integrity and provenance throughout its entire life-cycle. 

The architecture is designed according to NIST Zero Trust tenets \cite{stafford2020zero} where all data sources and computing services are considered resources, access to resources is determined by dynamic policy, authentication and authorization strictly enforced before access. The system separating the control plane (verification process) from the data plane (explanation generation). It consists of three logical components: the \textit{Untrusted Zone}, the \textit{Policy Enforcement Point (PEP)}, and the \textit{Policy Decision Point (PDP)}. Zero Trust treats auditor produced explanations such as SHAP objects untrusted by default, so instead of sending it directly to the user, it sends the explanation to \textit{PEP}. \textit{PEP} isolates the explainer through  micro-segmentation that ensures there is no direct path from explainer to the end user, and the explanation must go through \textit{PDP}. \textit{PEP} also sends an authorization request to \textit{PDP} with explainer object metadata. Policy Engine (PE) of \textit{PDP} evaluates the trust in the explainer object and then the trust decision is forwarded to Policy Administrator (PA), which issues authorization token, sends appropriate configuration to \textit{PEP} so that only trusted and verified explainer object is passed to the end user.  The Verifier ($\mathbf{V}$) acts as the immutable trust anchor by controlling the secure environment, managing cryptographic keys, and executing the non-repudiable verification protocols.

\subsection{XAI Pipeline: Roles, Assumptions, and Adversarial Goals}

The {\em ExplainGuard} framework is designed to counter a specific threat model involving a sophisticated, potentially malicious model auditor. We define the operational environment through the interaction of specific stakeholders and the technical constraints of the audit process.

\subsubsection{Stakeholder Roles and Trust Perimeters}
To establish a clear security boundary, we define the following entities:
\begin{itemize}
    \item \textbf{Model Developer ($\mathbf{Dev}$)} and \textbf{End User ($\mathbf{E}$)} are mutually trusted parties. $\mathbf{Dev}$ provides the original black-box model $f$ to $\mathbf{E}$, typically under a Machine Learning as a Service (MLaaS) paradigm.
    \item \textbf{Model Auditor ($\mathbf{A}$)} is an external, untrusted entity commissioned to evaluate $f$. The auditor is assumed to have black-box access to the model to generate post-hoc explanations using a model-agnostic explainer with limited time query access.
    \item \textbf{Verifier ($\mathbf{V}$)} represents the authoritative, trusted third-party service responsible for validating the integrity of the audit.
\end{itemize}

\subsubsection{Technical Assumptions}
The framework operates under the following technical constraints:
\begin{itemize}
    \item \textbf{Input and Task:} The audit is triggered by a sample input vector $x$. We focus on binary classification tasks where $f$ provides a deterministic prediction probability score for any given instance along with a class label.
    \item \textbf{Data and Tools:} For the scope of this work, we consider models trained on tabular data and Python coding environment.
    \item \textbf{Audit Motivation:} $\mathbf{E}$ (often a non-technical stakeholder) requires an audit for transparency, or $\mathbf{Dev}$ is regulated to provide black-box model access to auditor $\mathbf{A}$ to ensure an unbiased external evaluation.
\end{itemize}

\subsubsection{Adversarial Objectives} 
Auditor’s primary goal is \textit{Explanation Fairwashing}, meaning the generation of deceptive explanations ($\Phi_A$) to mislead both the $\mathbf{E}$ and the $\mathbf{V}$. We consider two primary attack vectors:
\begin{enumerate}
    \item \textbf{Model Substitution (Cloning):} $\mathbf{A}$ deploys an adversarial wrapper model ($f'$) that mimics the predictive behavior of $f$ but engineered to suppress sensitive features or output biased importance scores.
    \item \textbf{Explanation Forgery:} $\mathbf{A}$ submits feature importance values that are not derived from $f$. These may be randomized, manually manipulated, or generated from a corrupted explainer state to hide the model's true logic.
\end{enumerate}

\subsection{Operational Protocol and Key Management}
The operational protocol begins with the registration phase, where the model owner (either $\mathbf{E}$ or $\mathbf{Dev}$) registers the black-box model ($f$) with the $\mathbf{V}$, providing seed inputs to generate secret fingerprint inputs ($I_c = \{\mathcal{S}_{expl} \cup \mathcal{S}_{ood}$\}) via Algorithm \ref{OOD_data_generation_algo} (presented in \textit{Appendix C}) and establishing a cryptographic key pair ($K_{A_{private}}$ and $K_{A_{public}}$). During the authorization, $\mathbf{E}$ issues a unique, request-specific API key linked to model $f$ and a Request ID $R_{id}$ to the Auditor ($\mathbf{A}$), 
who is registered by the primary party and assigned their own cryptographic key pair ($K_{B_{private}}$ and $K_{B_{public}}$). Finally, in the submission phase, $\mathbf{A}$ submits a structured data package containing calculated attribution values ($\Phi_A = g(f',x)$) and the corresponding explainer object ($E_{obj}$); this package is serialized in a restricted, non-executable format and encrypted first with $K_{A_{public}}$ and then with $K_{B_{private}}$. This multi-layer encryption ensures non-repudiation by the auditor, as the signature is bound to the submission, while ensuring that only the model owner possesses the necessary keys to decrypt and access the final explanation result.

\vspace{-2mm}
\subsection{Architectural Components}
\subsubsection{The Untrusted Zone (The Auditor)}
Auditor ($\mathbf{A}$) is an external entity authorized to generate explanations for a specific query. Given that the model $f'$ is treated as a complete black-box with no access to internal gradients or layer architectures, $\mathbf{A}$ must rely on model-agnostic, post-hoc interpretation methods. We assume $\mathbf{A}$ operates in an adversarial environment and may attempt to submit manipulated explanations (Fairwashing) or malicious payloads (RCE attack). The Auditor generates a submission package $S_A = \{E_{obj}, \Phi_A, x, \mathcal{B}\}$, where $E_{obj}$ is the serialized post-hoc black-box explainer object (e.g., \texttt{shap.KernelExplainer} or \texttt{lime.lime\_tabular.LimeTabularExplainer}), $\Phi_A$ is the set of calculated feature attribution values, and $x$ is the input instance being explained and $\mathcal{B}$ is the background dataset used as reference distribution.

\subsubsection{Policy Enforcement Point (PEP)}
The PEP serves as the secure gateway between $\mathbf{A}$ (Untrusted) and $\mathbf{E}$. Zero-trust architecture ensures that no explanation can bypass \textit{PEP}. 
PEP has two primary responsibilities:
\textbf{(i) Traffic Interception \& Explainer Isolation:} \textit{PEP} intercepts $S_A$ and routes it to a secure, isolated container. This micro-segmentation isolates the explainer service so that it cannot communicate with any component except \textit{PEP}. By restricting all outbound traffic from the explainer, 
micro-segmentation guarantees that each explanation must pass through the \textit{PEP} and be evaluated by the \textit{PDP} before reaching the user. This prevents bypasses, limits lateral movement, and enforces a mandatory verification pipeline for explanation safety. \textbf{(ii) Authorization Builder:} 
    Another key objective of the \textit{PEP} is to construct and send a complete authorization request to the \textit{PDP} that includes metadata about the explainer output. It means that \textit{PEP} extracts model attributes such as model ID, explanation type, feature sensitivity, and user context, then packages these into a policy query for the \textit{PDP}. 

\subsubsection{Policy Decision Point (PDP)}
The \textit{PDP} is the ``Trust Engine'' of this architecture. It takes the sanitized artifacts from the \textit{PEP} and computes a trust score $\mathcal{T} \in \{0,1\}$ based on a hierarchical verification protocol. The \textit{PDP} has access to the {\em ground-truth store}, which contains the model ($f$) and the secret fingerprint set ($I_c$). 

\subsection{The Verification Protocol}

The \textit{PDP} enforces a ``Verify-Then-Trust'' policy through three distinct pillars. An explanation is only released if it satisfies all three pillars.

\subsubsection{Pillar I: Black-Box Asset Integrity}
This pillar verifies the identity and structural honesty of the model $f'$ using strictly input-output observations, defending against Model Substitution, Output Shuffling, and Explainer-Conditioned Scaffolding \cite{slack2020fooling} attacks.


\noindent \textbf{- Behavioral Fingerprinting:} The PDP validates the model identity by querying a secret set of inputs $I_c$ and verifying the output $f(I_c)$ against a pre-recorded ground truth $Y_{GT}$.
    \begin{equation}
        \mathcal{V}_{id} = \mathbb{I}\bigl(\| f(I_c) - f'(I_c) \| < \delta_{fin} \bigr)
    \end{equation}
    This test detects model substitution and unauthorized retraining, but does not by itself detect adversarial scaffolding.

\noindent\textbf{- Batch-Invariance Check:} To detect \textit{Output Shuffling} attacks where model predictions differ between individual and batch evaluation, we provide a simple and novel verification process:

    \begin{equation}
    \label{eq:output_shuffling_detect}
        \mathcal{V}_{\text{os}} = \mathbb{I}\left(
    \max_{i=1,\dots,n} \left| f'(I_{c_i}) - f'(I_c)[i] \right| < \delta_{os}
    \right)
    \end{equation}

    Honest model produces identical predictions regardless of evaluation mode, while output-shuffling models violate this invariance.

\noindent\textbf{- Explainer-Conditioned OOD Audit:} 
    To detect Slack-style adversarial scaffolding, the PDP probes whether the model conditions its internal logic on the distribution of explainer-generated inputs.

    For each seed input $x \in \mathcal{X}_{seed}$, the PDP constructs two perturbation distributions as provided in algorithm \ref{OOD_data_generation_algo} (see Appendix C):
    \begin{itemize}
        \item $S_{expl}(x)$: explainer-style perturbations (e.g., SHAP/LIME masking and jittering)
        \item $S_{ood}(x)$: non-explainer OOD perturbations (e.g., feature swapping and Gaussian noise)
    \end{itemize}

    The PDP computes the distributional discrepancy:
    \[
        \Delta(x) =
        \left\|
        \mathbb{E}_{z \sim S_{expl}(x)}[f'(z)]
        -
        \mathbb{E}_{z \sim S_{ood}(x)}[f'(z)]
        \right\|.
    \]

    The scaffold detection statistic is then
    \begin{equation}
    \label{eq:ood_scaffold_detect}
        \mathcal{V}_{sc} =
        \mathbb{I}\left(
        \frac{1}{|\mathcal{X}_{seed}|}
        \sum_{x \in \mathcal{X}_{seed}}
        \Delta(x)
        > \delta_{ood}
        \right).
    \end{equation}

    An honest model, whose logic does not depend on the explainer’s sampling distribution, satisfies $\Delta(x) \approx 0$.  
    A scaffolded model exhibits a systematic discrepancy because it activates different internal predictors on $S_{expl}$ and $S_{ood}$. This protocol extends the conditional anomaly detection framework proposed by Carmichael and Scheirer \cite{carmichael2023unfooling} by introducing a contrastive OOD baseline. While the original framework focuses on identifying explainer-specific perturbations ($\mathcal{S}_{expl}$), our algorithm generates a neutral control set ($\mathcal{S}_{ood}$) using group swapping and Gaussian noise. This allow the verifier to differentiate between a model that is naturally unstable on off-manifold data and one that is intentionally scaffolded to deceive an auditor's specific sampling manifold.

\noindent\textbf{- Biased Sampling Detection:}
    To detect whether the background distribution $\mathcal{B}$ has been adversarially biased, we jointly test for both \textit{distributional shift} and \textit{mean shift} between model outputs on the seed data $\mathcal{X}_{\text{seed}}$ and on the background samples $\mathcal{B}$, following the two-stage fraud detection logic by  LaBerge et al. \cite{laberge2022fool_biased_sampling}. Let $
    \mathcal{Y}_{\text{seed}} = \{ f'(x) : x \in X_{\text{seed}} \}, 
    \quad
    \mathcal{Y}_{\mathcal{B}} = \{ f'(b) : b \in \mathcal{B} \}
    $. Then a two-sample Kolmogorov--Smirnov (KS) test to compare the empirical cumulative distributions of $\mathcal{Y}_{\text{seed}}$ and $\mathcal{Y}_{\mathcal{B}}$, and a Wald test to detect shifts in their respective sample means is done.

    {Verification Rule:}
    Let $p_{\text{ks}}$ and $p_{\text{wald}}$ be the corresponding two-sided p-values from these tests. The verification is defined as:
    \begin{equation}
    \mathcal{V}_{\text{sm}}
    =
    \mathbb{I}
    \left(
    p_{\text{ks}} \ge \frac{\delta_{sample}}{2}
    \;\wedge\;
    p_{\text{wald}} \ge \frac{\delta_{sample}}{2}
    \right)
    \label{eq:biased-sampling-hybrid}
    \end{equation}
    where $\delta_{sample}$ is a user-defined significance level. A rejection indicates that the background distribution induces a statistically significant shift in model outputs and is flagged as potentially biased.


\subsubsection{Pillar II: Semantic Validity}
This policy verifies that the explanation is mathematically consistent with the model's output without requiring access to internal gradients. We enforce two axioms that form the theoretical foundation of additive attribution methods such as \textsc{KernelSHAP} and \textsc{LIME}:

    \noindent\textbf{- Efficiency (Local Accuracy):} For a faithful explanation, the sum of all feature attributions plus the baseline must equal the model's prediction. For \textsc{SHAP}, this is $\phi_0 + \sum \phi_i = f(x)$ where $\phi_0 = \mathbb{E}[f(X)]$. For \textsc{LIME}, this is $\phi_0 + \sum \phi_i x_i \approx f(x)$ where $\phi_0$ is the local intercept \cite{fryer2021shapley}. The Null-Player Axiom requires that if a feature $i$ has a negligible marginal contribution, defined as $\sigma_i < \delta'$ where $\sigma_i$ is defined as the maximum marginal contribution of feature $i$, then its corresponding attribution $\phi_i$ must also be negligible \cite{fryer2021shapley}, such that $|\phi_i| < \delta_2$.
    
    \noindent \textbf{- Null-Player Axiom:} If perturbing a feature does not change the model output, its attribution must be zero. This holds for both \textsc{SHAP} (via Shapley value properties) and \textsc{LIME} (via local linear model weights).

\begin{equation}
\begin{aligned}
\mathcal{V}_{\text{ax}} = &\ \mathbb{I}\left(\left| \phi_0 + \sum_{i=1}^{d} \psi_i - f'(x) \right| < \delta_1 \right) \\
& \wedge \ \mathbb{I}\left( \max_{i: \sigma_i < \delta'} |\phi_i| < \delta_2 \right)
\end{aligned}
\end{equation}

This check ensures that the Auditor has accounted for the entire prediction shift and has not assigned importance to inactive features, preventing both omission and fabrication of causal effects.

\subsubsection{Pillar III: Feature Faithfulness Verification}

To verify that the Auditor's top-$k$ attributions are not arbitrary but accurately reflect the model's causal logic, we implement a comparative deletion-based evaluation motivated from the work of Wei et.al. \cite{xNIDS_285389} and Samek et.al. \cite{samek2016evaluating}. This protocol verifies that the features identified as most influential by the Auditor ($A$) cause a significantly greater impact on model behavior than a random baseline.

    \noindent \textbf{- Feature Ranking:} Rank features by descending absolute attribution magnitude:
    \[
    i_1, i_2, \dots, i_d \quad \text{such that} \quad |\phi_{i_1}| \ge |\phi_{i_2}| \ge \dots \ge |\phi_{i_d}|
    \]

    \noindent \textbf{- Top-$k$ Ablation Protocol:} For $k = 1, \dots K$, generate ablated instances $x^{(k)}$ by replacing the top-$k$ features with values $b_i$ drawn from the validated background distribution $\mathcal{B}$. 
    
    \noindent \textbf{- Directional Degradation:} To account for signed attributions (SHAP/LIME), we measure the \textit{magnitude of change} in model output relative to the original prediction $y_0 = f(x)$:
    \[
    \Delta_k = |f'(x) - f'(x^{(k)})|
    \]

    \noindent \textbf{- Faithfulness AUC ($\text{AUC}_{A}$):} We calculate the normalized Area Under the Deletion Curve for the Auditor's ranking:
    \[
    \text{AUC}_{A} = \frac{1}{K} \sum_{k=1}^{K} \Delta_k
    \]

    \noindent \textbf{- Random Baseline AUC ($\text{AUC}_{M}$):} We generate $M$ random feature permutations $\pi_1, \dots, \pi_M$. Let $\bar{\Delta}_j = \frac{1}{K}\sum_{k=1}^{K} |f'(x) - f'(x^{(\pi_j,k)})|$ denote the per-permutation deletion score accross $K$ ablation steps then:
    \[
    \text{AUC}_{M} = \frac{1}{M} \sum_{j=1}^{M} \left( \bar\Delta_j\right)
    \]


    \noindent \textbf{- Faithfulness Gap:} The discriminative margin between the Auditor's ranking and the random baseline:
\[
\Delta_{\text{fd}} = \text{AUC}_{A} - \text{AUC}_{M}
\]

\noindent \textbf{- Adaptive Threshold:} Let $\sigma_A^2$ and $\sigma_M^2$ denote the sample variances of the per-instance Auditor scores $\{\text{AUC}_A^{(i)}\}_{i=1}^{N}$ and the pooled permutation scores $\{\bar\Delta_j\}_{j=1}^{n}$ ($n=N\times M$), respectively.
The significance threshold is then the pooled standard error:
\[
\delta_{M} = \sqrt{\frac{\sigma_A^2}{N} + \frac{\sigma_M^2}{n}}
\]

\noindent \textbf{- Verification Rule:} An explanation is faithful if $\Delta_{\text{fd}}$ exceeds this uncertainty by a significant margin:
\begin{equation}
    \mathcal{V}_{\text{fd}} = \mathbb{I}\left( \Delta_{\text{fd}} \ge z \cdot \delta_{M} \right)
\end{equation}
where $z$ is a significance coefficient (e.g., $z=1.96$ for 95\% confidence). The threshold $\delta_M$ is self-adaptive; thus, we do not need to specify it explicitly. A failure indicates that the Auditor's top-ranked features are statistically indistinguishable from noise. This effectively detects ``Fairwashing" where an Auditor assigns high importance to irrelevant features to distract from biased ones.


\section{ExplainGuard Compliance and Logical Proof of Verification}
\label{sec:logic_proof}

\subsection{Compliance with NIST Zero Trust Standards}
ExplainGuard is designed not merely as a theoretical security framework but as a functional implementation of the NIST SP 800-207 standard. We map the foundational tenets of Zero Trust directly to our architectural components to ensure the explainability supply chain remains resilient against adversarial manipulation. We address Tenet 3 (Per-Session Access) through the use of ephemeral, micro-segmented containers at the Policy Enforcement Point (PEP), ensuring that trust is never persistent across explanation submissions and that the explainer object ($E_{obj}$) is strictly isolated from the end-user until verification is complete. Furthermore, we satisfy Tenet 4 (Dynamic Policy) by employing a multi-factor Policy Decision Point (PDP) that evaluates the trust score $\mathcal{T}$ based on real-time evidence across three layers of verification. The PDP first executes a Black-Box Asset Integrity (Pillar I) audit to evaluate the observable state and identity of the black-box model and the statistical representativeness of the background data $\mathcal{B}$. It then enforces Semantic Validity (Pillar II) to ensure the mathematical consistency of the attribution values $\Phi_A$ relative to the model's output $f'(x)$. Finally, it validates the explanation through Feature Faithfulness Verification (Pillar III) by verifying that the top-$k$ rankings induce a significantly higher cumulative impact on model degradation than a random baseline. This holistic approach ensures that security decisions are driven by real-time, data-centric risk assessment rather than static network perimeters or implicit auditor trust.

\vspace{1mm}
\subsection{Trust Decision and Feedback}
The final Trust Decision $\mathcal{T}$ is computed as the logical conjunction of the verification outcomes from all three architectural pillars. This strict ``Fail-Closed" logic ensures that any deviation in model identity, mathematical consistency, or feature faithfulness results in a total rejection of the explanation. The trust score is defined as:
\begin{equation}
    \mathcal{T} = \mathcal{V}_{id} \land \mathcal{V}_{os} \land \mathcal{V}_{sc} \land \mathcal{V}_{sm} \land \mathcal{V}_{ax} \land \mathcal{V}_{fd}
\end{equation}
where $\mathcal{V}_{id}$, $\mathcal{V}_{os}$. $\mathcal{V}_{sc}$ and $\mathcal{V}_{sm}$ represent the structural and distributional integrity checks of Pillar I, $\mathcal{V}_{ax}$ denotes the semantic validity of Pillar II, and $\mathcal{V}_{fd}$ represents the feature rank correlation derived in Pillar III.

If $\mathcal{T} = 1$, the \textit{PDP} issues a success token to the PEP, which then releases the decrypted, verified explanation artifact to the End User. Conversely, if $\mathcal{T} = 0$, the PEP immediately drops the submission and terminates the session.




\section{Experimental Case Study and Evaluation}
\label{sec:experiments}
To validate the ExplainGuard framework, we conducted a rigorous empirical evaluation using the recidivism risk prediction task on the COMPAS dataset \cite{COMPAS_larson2016we} with 8 features. The dataset was preprocessed and partitioned into training, testing, and audit subsets. An XGBoost classifier was trained on the numerical feature matrix, yielding an $F_1$ score of 0.65, and serialized into a standardized JSON artifact. To prevent direct model inspection, a standardized black-box wrapper endpoint was established for the auditor. Both KernelSHAP and LIME submission packages were augmented with self-contained, pickle-safe model prediction wrappers ($\text{Explainer.Model.f}$ embedded; available natively for KernelSHAP, and explicitly enforced for $\text{LimeTabularExplainer}$). All experiments are executed locally, leaving cryptographic foundations for a production-level implementation.

\subsection{Comparative Audit Results Across Threat Models}
We evaluated the clean baseline model alongside four sophisticated adversarial attack scenarios:
\begin{itemize}
    \item \textbf{Scenario 1: Clean Baseline:} Standard random sampling of background data using random seed 42 with default KernelSHAP and LIME attributions.
    \item \textbf{Scenario 2: Scaffolding OOD Attack:} Alters model outputs on out-of-distribution explainer probes to manipulate global importances. Target features include \texttt{age}, \texttt{sex}, and \texttt{race\_African-American}, using 0 as the random seed for background data.
    \item \textbf{Scenario 3: Output Shuffling Attack:} Exploits batch-level rank permutations to obfuscate feature attributions. Target features include \texttt{sex} and \texttt{race\_African-American}, using 100 as the random seed for background data.
    \item \textbf{Scenario 4: Biased Sampling Attack:} Optimizes weighted background datasets via convex optimization to suppress protected feature attributions (target feature \texttt{sex}) while striving to remain beneath statistical detection thresholds.
    \item \textbf{Scenario 5: Random Baseline:} A random control case where SHAP and LIME attributions are generated randomly with randomized feature ranks, using 1334 as the random seed for background data.
\end{itemize}

Table~\ref{tab:pillar_audit_results} summarizes the performance across all verification checks spanning ExplainGuard's three security pillars, displaying maximum discrepancies, empirical $p$-values, delta thresholds, and final pass/fail outcomes showing the efficacy in identifying all adversarial scenarios.  The complete implementation can be found in this Github repository : \url{https://github.com/MarazMia/ExplainGuard}.

\subsection{Key Insights}
\begin{itemize}
    \item \textbf{Output Shuffling Detection ($V_{id}, V_{os}$):} Elevated metric values ($0.105244$) exceeding the threshold successfully capture the output shuffling attacks, as reported in Table~\ref{tab:pillar_audit_results}.
    \item \textbf{Scaffolding OOD Detection ($V_{sc}$):} Identifies out-of-distribution branch-switching behavior in scaffolding probes, flagging discrepancies of $0.111597$ against the $< 0.080$ threshold limit.
    \item \textbf{Biased Sampling Defense ($V_{\text{sm}}$):} While optimization techniques \cite{laberge2022fool_biased_sampling} bypass lenient baseline thresholds ($p \ge 0.025$), hardening ExplainGuard's verification harness to a stricter margin ($\ge 0.20$) successfully traps warped background distributions ($p_{\text{KS}} = 0.1124$, $p_{\text{Wald}} = 0.1614$), resulting in robust security rejections.
    \item \textbf{Random Submission Defense ($V_{\text{ax}}, V_{\text{fd}}$):} Both verification checks successfully catch randomized feature attributions submitted by the auditor, evaluated via axiom thresholds ($1.137555$ for SHAP and $1.454038$ for LIME) and adaptive feature fidelity thresholds ($0.006$ for SHAP and $0.0106$ for LIME), which dynamically scale using internal statistical importance values ($z$) and standard errors.
\end{itemize}


\begin{table*}[!h]
\centering
\vspace{2mm}
\caption[ExplainGuard Pillar I--III Security Verification Results]{ExplainGuard Pillar I--III Security Verification Results Across Attack Scenarios for COMPAS Dataset and XGBoost Model. ($\smallhalfleft$ indicates applicable for both LIME and SHAP, $\smallfullsolid$ for SHAP only, and $\smallunfilled$ for LIME only.)}
\label{tab:pillar_audit_results}
\small
\resizebox{\textwidth}{!}{%
\begin{tabular}{lcccccc}
\hline
\textbf{Verification Check} & \textbf{Threshold} & \textbf{Clean Baseline} & \textbf{OOD Scaffolding} & \textbf{Output Shuffling} & \textbf{Biased Sampling} & \textbf{Random} \\ \hline
$V_{id}$ (\smallhalfleft) & $\delta_{fin} = 0.005 >$ & $0.0$ & $0.0$ & $0.105244^*$ & $0.0$ & $0.0$ \\
$V_{os}$ (\smallhalfleft) & $\delta_{os} = 0.005 >$ & $0.0$ & $0.0$ & $0.105244^*$ & $0.0$ & $0.0$ \\
$V_{sc}$ (\smallhalfleft) & $\delta_{ood} = 0.08 > $ & $0.057689$ & $0.111597^*$ & $0.057689$ & $0.057689$ & $0.057689$ \\
$V_{sm}$ ($p_{\text{KS}}$) (\smallhalfleft) & $\frac{\delta_{sm}}{2} = 0.2 < $ & $0.3959$ & $0.8693$ & $0.5487$ & $\textbf{$0.1124^*$}$ & $0.5487$ \\
$V_{sm}$ ($p_{\text{Wald}}$) (\smallhalfleft) & $\frac{\delta_{sm}}{2} = 0.2 < $ & $0.262$ & $0.8587$ & $0.7878$ & $\textbf{$0.1614^*$}$ & $0.4409$ \\
$V_{ax}$ (\smallfullsolid) & $\delta_{1} = 0.05 > $ & $4.08 \times 10^{-8}$ & $1.11 \times 10^{-16}$ & $2.22 \times 10^{-16}$ & $4.53 \times 10^{-8}$ & $1.137555^*$ \\
$V_{ax}$ (\smallfullsolid) & $\delta_{2} = 0.05 > $ & $0.0$ & $0.0$ & $0.0610$ & $0.1650$ & $0.0$ \\
$V_{ax}$ (\smallunfilled) & $\delta_{1} = 0.4 > $ & $0.26604$ & $0.330753$ & $0.370767$ & $N/A$ & $1.454038^*$ \\
$V_{ax}$ (\smallunfilled) & $\delta_{2} = 0.05 > $ & $0.0$ & $0.0$ & $0.0$ & $N/A$ & $0.0$ \\
$V_{fd}$ (\smallfullsolid) & $\delta_M = \text{adpt.} > $ & $0.0127$ & $0.0132$ & $0.0166$ & $0.019$ & $0.006^*$ \\
$V_{fd}$ (\smallunfilled) & $\delta_M = \text{adpt.} > $ & $0.0121$ & $0.0115$ & $0.0147$ & $N/A$ & $0.0106^*$ \\
$\mathcal{T}$ (\smallhalfleft) & $\text{pass/fail}$ & \textbf{pass} & \textbf{fail} & \textbf{fail} & \textbf{fail} & \textbf{fail} \\ \hline
\end{tabular}%
}
\vspace{2mm}
\footnotesize
$(^*)$ indicating the failed cases in verification check. 
\end{table*}

\section{Discussion, Limitations and Future Directions}
\label{sec:limitations}
To the best of our knowledge, this is the first work that proposes to systematically verifies XAI results produced by an external auditor with a zero-trust architecture. This design renders the framework suitable for deployment in cloud or server-based environments where neither the model developer nor the auditor is assumed to be fully trustworthy.

A natural concern is the scenario where the original model $f$, supplied by the $\mathbf{Dev}$, is itself a wrapper-based adversarial (scaffolded) model. In this case, detection remains feasible using only black-box access: based on the seed dataset $\mathcal{X}_{\text{seed}}$, and the verification rules in Eq.~\ref{eq:output_shuffling_detect} and Eq.~\ref{eq:ood_scaffold_detect}, the Batch-Invariance Check and OOD Stability audit can both flag anomalous behavior as these tests operate solely on input–output inconsistencies of a single black-box model. For selecting the top-$k$ features, using the heuristic $k = \sqrt{|F|}$, where $|F|$ is the number of input features, is recommended as a standard rule of thumb. For the LIME $V_{ax}$ threshold $\delta_1$, a more lenient value of $0.4$ is adopted because it depends on the original model's performance. Since LIME relies on an internal surrogate linear model as an approximation, and the underlying COMPAS model exhibits moderate performance ($F_1$ score of $0.65$), it inherently suffers from approximation noise. Furthermore, diverse background distributions were evaluated across multiple random seeds to ensure that the established thresholds avoid random guessing and generalize robustly across diverse test cases.

In this paper, our scope is limited to tabular data and primarily centered on SHAP and LIME while broader cross-explainer generalization remains an open challenge. Also, extending {\em ExplainGuard} to other data modalities and non-additive or gradient-based explanation methods (e.g., Integrated Gradients, Grad-CAM, counterfactual explainers) still remains as open problem. The threat model focuses on two representative wrapper-based attacks, namely the scaffolding and output shuffling, under a black-box setting and does not exhaust the space of possible explanation forgery strategies, nor does it cover white-box attacks, inherently interpretable models, or non-additive explainers. Due to the black-box assumption, we also do not consider orthogonal model-level threats such as back-doors, direct weight manipulation, data poisoning, or membership inference, which are complementary to the explanation-integrity problem studied here. The robustness of the detection mechanisms further depends on the representativeness of the seed dataset $\mathcal{X}_{\text{seed}}$, and inadequate coverage may allow adversarial evasion. Finally, while the performance overhead is acceptable for many use cases, it can be optimized further, and we plan to formalize verification guarantees, extend the framework to defend against privacy-leakage issues via explanations, and address system-level security concerns such as secure logging and resilience to denial-of-service attacks.

\section{Conclusion}
\label{sec:conclusion}

In summary, this paper exposes a critical vulnerability in modern XAI pipelines: the lack of security guarantees for post-hoc explanations. We show that wrapper-based attacks such as fairwashing, biased sampling, and explanation scaffolding can compromise explanation trustworthiness without modifying the underlying model. To mitigate these risks, we introduce {\em ExplainGuard}, a zero trust architecture–based framework that enforces per-session access control, dynamic policy verification, and continuous behavioral monitoring for maintaining explanation integrity before it is presented to end-users. Although our framework is proposed for some specific XAI methods, it has the potential to be extended further in addition with other robust and mathematically sound pillars.

\section*{Acknowledgment}
This work is partially supported and developed under the National Science Foundation Research Traineeship (NRT) award \# 2346001, and NSF grants \# 2416990, \#2230609 at Tennessee Tech University.

\bibliographystyle{IEEEtran}
\bibliography{reference}

\begin{thebibliography}{10}
\providecommand{\url}[1]{#1}
\csname url@samestyle\endcsname
\providecommand{\newblock}{\relax}
\providecommand{\bibinfo}[2]{#2}
\providecommand{\BIBentrySTDinterwordspacing}{\spaceskip=0pt\relax}
\providecommand{\BIBentryALTinterwordstretchfactor}{4}
\providecommand{\BIBentryALTinterwordspacing}{\spaceskip=\fontdimen2\font plus
\BIBentryALTinterwordstretchfactor\fontdimen3\font minus \fontdimen4\font\relax}
\providecommand{\BIBforeignlanguage}[2]{{%
\expandafter\ifx\csname l@#1\endcsname\relax
\typeout{** WARNING: IEEEtran.bst: No hyphenation pattern has been}%
\typeout{** loaded for the language `#1'. Using the pattern for}%
\typeout{** the default language instead.}%
\else
\language=\csname l@#1\endcsname
\fi
#2}}
\providecommand{\BIBdecl}{\relax}
\BIBdecl

\bibitem{hosain2024explainable}
M.~T. Hosain, J.~R. Jim, M.~Mridha, and M.~M. Kabir, ``Explainable ai approaches in deep learning: Advancements, applications and challenges,'' \emph{Computers and electrical engineering}, vol. 117, p. 109246, 2024.

\bibitem{singh2024towards}
J.~Singh, S.~Rani, and G.~Srilakshmi, ``Towards explainable ai: interpretable models for complex decision-making,'' in \emph{2024 International Conference on Knowledge Engineering and Communication Systems (ICKECS)}, vol.~1.\hskip 1em plus 0.5em minus 0.4em\relax IEEE, 2024, pp. 1--5.

\bibitem{shap_original_lundberg2017unified}
S.~M. Lundberg and S.-I. Lee, ``A unified approach to interpreting model predictions,'' \emph{Advances in neural information processing systems}, vol.~30, 2017.

\bibitem{lime_original_ribeiro2016should}
M.~T. Ribeiro, S.~Singh, and C.~Guestrin, ``" why should i trust you?" explaining the predictions of any classifier,'' in \emph{Proceedings of the 22nd ACM SIGKDD international conference on knowledge discovery and data mining}, 2016, pp. 1135--1144.

\bibitem{IG_original_sundararajan2017axiomatic}
M.~Sundararajan, A.~Taly, and Q.~Yan, ``Axiomatic attribution for deep networks,'' in \emph{International conference on machine learning}.\hskip 1em plus 0.5em minus 0.4em\relax PMLR, 2017, pp. 3319--3328.

\bibitem{selvaraju2017grad_cam_original}
R.~R. Selvaraju, M.~Cogswell, A.~Das, R.~Vedantam, D.~Parikh, and D.~Batra, ``Grad-cam: Visual explanations from deep networks via gradient-based localization,'' in \emph{Proceedings of the IEEE international conference on computer vision}, 2017, pp. 618--626.

\bibitem{binder2016layer_LRP_original}
A.~Binder, G.~Montavon, S.~Lapuschkin, K.-R. M{\"u}ller, and W.~Samek, ``Layer-wise relevance propagation for neural networks with local renormalization layers,'' in \emph{International conference on artificial neural networks}.\hskip 1em plus 0.5em minus 0.4em\relax Springer, 2016, pp. 63--71.

\bibitem{shrikumar2017learning_deep_lift_original}
A.~Shrikumar, P.~Greenside, and A.~Kundaje, ``Learning important features through propagating activation differences,'' in \emph{International conference on machine learning}.\hskip 1em plus 0.5em minus 0.4em\relax PMlR, 2017, pp. 3145--3153.

\bibitem{verma2024counterfactual}
S.~Verma, V.~Boonsanong, M.~Hoang, K.~Hines, J.~Dickerson, and C.~Shah, ``Counterfactual explanations and algorithmic recourses for machine learning: A review,'' \emph{ACM Computing Surveys}, vol.~56, no.~12, pp. 1--42, 2024.

\bibitem{pachl2025view}
E.~Pachl, F.~Langer, T.~Markert, and J.~M. Lorenz, ``A view on vulnerabilites: The security challenges of xai (academic track),'' in \emph{Symposium on Scaling AI Assessments (SAIA 2024)}.\hskip 1em plus 0.5em minus 0.4em\relax Schloss Dagstuhl--Leibniz-Zentrum f{\"u}r Informatik, 2025, pp. 12--1.

\bibitem{accuracy_and_xai_papenmier_2022}
\BIBentryALTinterwordspacing
A.~Papenmeier, D.~Kern, G.~Englebienne, and C.~Seifert, ``It’s complicated: The relationship between user trust, model accuracy and explanations in ai,'' \emph{ACM Trans. Comput.-Hum. Interact.}, vol.~29, no.~4, Mar. 2022. [Online]. Available: \url{https://doi.org/10.1145/3495013}
\BIBentrySTDinterwordspacing

\bibitem{slack2020fooling}
D.~Slack, S.~Hilgard, E.~Jia, S.~Singh, and H.~Lakkaraju, ``Fooling lime and shap: Adversarial attacks on post hoc explanation methods,'' in \emph{Proceedings of the AAAI/ACM Conference on AI, Ethics, and Society}, 2020, pp. 180--186.

\bibitem{outputyuan2024fooling}
\BIBentryALTinterwordspacing
J.~Yuan and A.~Dasgupta, ``Fooling shap with output shuffling attacks,'' 2024. [Online]. Available: \url{https://arxiv.org/abs/2408.06509}
\BIBentrySTDinterwordspacing

\bibitem{mia2025explainable}
M.~Mia and M.~M.~A. Pritom, ``Explainable but vulnerable: Adversarial attacks on xai explanation in cybersecurity applications,'' \emph{arXiv preprint arXiv:2510.03623}, 2025.

\bibitem{noppel2024sok}
M.~Noppel and C.~Wressnegger, ``Sok: Explainable machine learning in adversarial environments,'' in \emph{2024 IEEE Symposium on Security and Privacy (SP)}.\hskip 1em plus 0.5em minus 0.4em\relax IEEE, 2024, pp. 2441--2459.

\bibitem{aivodji2019fairwashing}
U.~A{\"\i}vodji, H.~Arai, O.~Fortineau, S.~Gambs, S.~Hara, and A.~Tapp, ``Fairwashing: the risk of rationalization,'' in \emph{International Conference on Machine Learning}.\hskip 1em plus 0.5em minus 0.4em\relax PMLR, 2019, pp. 161--170.

\bibitem{stafford2020zero}
V.~Stafford, ``Zero trust architecture,'' \emph{NIST special publication}, vol. 800, no. 207, pp. 800--207, 2020.

\bibitem{zerotrustinmlops}
H.~N. V. S.~M. Krishna~Tungala, G.~Yeleswarapu, M.~Shivnatri, and S.~K. Irujolla, ``A zero trust framework with ai-driven identity and intrusion detection for multi-cloud mlops,'' in \emph{2025 13th International Symposium on Digital Forensics and Security (ISDFS)}, 2025, pp. 1--6.

\bibitem{kuppa2020blackbox}
A.~Kuppa and N.-A. Le-Khac, ``Black box attacks on explainable artificial intelligence(xai) methods in cyber security,'' in \emph{2020 International Joint Conference on Neural Networks (IJCNN)}, 2020, pp. 1--8.

\bibitem{yuan2023a}
\BIBentryALTinterwordspacing
J.~Yuan and A.~Dasgupta, ``A simple scoring function to fool {SHAP}: Stealing from the one above,'' in \emph{XAI in Action: Past, Present, and Future Applications}, 2023. [Online]. Available: \url{https://openreview.net/forum?id=iMR4ukkUFU}
\BIBentrySTDinterwordspacing

\bibitem{carmichael2023unfooling}
Z.~Carmichael and W.~J. Scheirer, ``Unfooling perturbation-based post hoc explainers,'' in \emph{Proceedings of the AAAI conference on artificial intelligence}, vol.~37, no.~6, 2023, pp. 6925--6934.

\bibitem{laberge2022fool_biased_sampling}
G.~Laberge, U.~A{\"\i}vodji, S.~Hara, F.~Khomh \emph{et~al.}, ``Fool shap with stealthily biased sampling,'' \emph{arXiv preprint arXiv:2205.15419}, 2022.

\bibitem{fryer2021shapley}
D.~Fryer, I.~Str{\"u}mke, and H.~Nguyen, ``Shapley values for feature selection: The good, the bad, and the axioms,'' \emph{Ieee Access}, vol.~9, pp. 144\,352--144\,360, 2021.

\bibitem{xNIDS_285389}
\BIBentryALTinterwordspacing
F.~Wei, H.~Li, Z.~Zhao, and H.~Hu, ``{xNIDS}: Explaining deep learning-based network intrusion detection systems for active intrusion responses,'' in \emph{32nd USENIX Security Symposium (USENIX Security 23)}.\hskip 1em plus 0.5em minus 0.4em\relax Anaheim, CA: USENIX Association, Aug. 2023, pp. 4337--4354. [Online]. Available: \url{https://www.usenix.org/conference/usenixsecurity23/presentation/wei-feng}
\BIBentrySTDinterwordspacing

\bibitem{samek2016evaluating}
W.~Samek, A.~Binder, G.~Montavon, S.~Lapuschkin, and K.-R. M{\"u}ller, ``Evaluating the visualization of what a deep neural network has learned,'' \emph{IEEE transactions on neural networks and learning systems}, vol.~28, no.~11, pp. 2660--2673, 2016.

\bibitem{COMPAS_larson2016we}
J.~Larson, S.~Mattu, L.~Kirchner, and J.~Angwin, ``How we analyzed the compas recidivism algorithm,'' \emph{ProPublica (5 2016)}, vol.~9, no.~1, pp. 3--3, 2016.

\bibitem{verma2023insecure}
A.~Verma, ``Insecure deserialization detection in python,'' 2023.

\end{thebibliography}

\appendix
\subsection{Need for an Explainer Object}
\label{sec:manipulation_scenario}
ExplainGuard mandates the submission of the original serialized explainer object ($E_{obj}$) utilized by the auditor to generate the feature attribution values. This requirement is fundamental to the Zero Trust model, as the $E_{obj}$ serves as the technical evidence of the specific model manifold the auditor probed during the explanation process. Without this object, the verifier possesses no mechanism to authenticate whether the reported attributions $\Phi_A$ were derived from the legitimate black-box model $f$ or a manipulated surrogate $f'$. By enforcing the submission of $E_{obj}$, the architecture allows the PDP to detect four distinct configurations. In the ideal case, the Auditor provides an $E_{obj}$ correctly applied to $f$, with $\Phi_A$ faithfully calculated for $f$. Conversely, a malicious Auditor may attempt to submit an $E_{obj}$ applied to a surrogate $f'$ while providing pre-calculated attributions for $f$ to mask the deception. A more common attack vector involves applying the explainer to the legitimate model $f$ while fabricating the importance values $\Phi_A$ to hide bias, or the most aggressive approach, where both the explainer object and the attributions are derived from a scaffolded surrogate $f'$. By intercepting the $E_{obj}$ at the Policy Enforcement Point, the Verifier can re-execute queries against the internal model reference preserved within the object, ensuring that the explanation artifact is intrinsically bound to the actual decision logic of the registered model $f$.

\vspace*{-2mm}

\begin{algorithm}[!b]
\caption{Black-Box Model Fingerprinting data generation via Deterministic--Stochastic OOD Probing}
\label{OOD_data_generation_algo}
\KwIn{Seed set $\mathcal{X}_{seed}=\{\mathbf{x}_1,\dots,\mathbf{x}_n\}$, model $f$, sparsity $k$, samples per mode $M$}
\KwOut{Explainer fingerprints $\mathcal{S}_{expl}$, OOD fingerprints $\mathcal{S}_{ood}$}

$\mathcal{S}_{expl} \leftarrow \emptyset, \quad \mathcal{S}_{ood} \leftarrow \emptyset$\;

\ForEach{$\mathbf{x} \in \mathcal{X}$}{

  \tcp{Explainer-style perturbations}
  \For{$m=1$ \KwTo $M$}{
    $\mathbf{z} \leftarrow \mathbf{x}$\;
    Draw $s \sim \{1,2\}$ uniformly\;

    \eIf{$s=1$}{
        \tcp{SHAP-style masking}
        Choose $(d-k)$ indices $I \subset \{1,\ldots,d\}$\;
        \ForEach{$i \in I$}{
            $\mathbf{z}^{(i)} \leftarrow \mathrm{median}\{\mathbf{x}_j^{(i)} : \mathbf{x}_j \in \mathcal{X}\}$\;
        }
    }{
        \tcp{LIME-style jittering}
        Choose $k$ indices $I \subset \{1,\ldots,d\}$\;
        \ForEach{$i \in I$}{
            Draw $\mathbf{z}^{(i)}$ uniformly from 
            $\{\mathbf{x}_j^{(i)} : \mathbf{x}_j \in \mathcal{X}\setminus\{\mathbf{x}\}\}$\;
        }
    }

    $\mathcal{S}_{expl} \leftarrow \mathcal{S}_{expl} \cup \{(\mathbf{z},f(\mathbf{z}))\}$\;
  }

  \tcp{Non-explainer OOD perturbations}
  \For{$m=1$ \KwTo $M$}{
    $\mathbf{z} \leftarrow \mathbf{x}$\;
    Draw $s \sim \{3,4\}$ uniformly\;

    \eIf{$s=3$}{
        \tcp{Group feature swapping}
        Choose a subset of indices $G \subset \{1,\ldots,d\}$\;
        Draw a reference point $\mathbf{x}_{ref}$ from $\mathcal{X}\setminus\{\mathbf{x}\}$\;
        \ForEach{$i \in G$}{
            $\mathbf{z}^{(i)} \leftarrow \mathbf{x}_{ref}^{(i)}$\;
        }
    }{
        \tcp{Gaussian feature noise}
        Choose $k$ indices $I \subset \{1,\ldots,d\}$\;
        \ForEach{$i \in I$}{
            $\mathbf{z}^{(i)} \leftarrow \mathbf{x}^{(i)} + \mathcal{N}(0,\mathrm{Var}(\mathcal{X}^{(i)}))$\;
        }
    }

    $\mathcal{S}_{ood} \leftarrow \mathcal{S}_{ood} \cup \{(\mathbf{z},f(\mathbf{z}))\}$\;
  }

}
\Return $\mathcal{S}_{expl}, \mathcal{S}_{ood}$\;
\end{algorithm}

\subsection{Defensive Serialization and Object Sandboxing}
To neutralize the inherent risks of Remote Code Execution (RCE) and object injection attacks, ExplainGuard enforces a strict prohibition on executable serialization formats. Standard Python serialization via the \texttt{pickle} module is fundamentally incompatible with Zero Trust principles, as it allows for the construction of malicious byte-streams that execute arbitrary code upon deserialization \cite{verma2023insecure}. Given that RCE grants an adversary lateral movement and full administrative control, its use for untrusted external inputs is strictly forbidden \cite{verma2024counterfactual}. Consequently, the entire lifecycle of the explainer object ($E_{obj}$), from ingestion to structural inspection, is confined within an ephemeral, non-privileged sandbox at the Policy Enforcement Point (PEP). This micro-segmented environment ensures that any potential exploitation attempt is contained, preventing escape to the wider system or the Model Owner's internal assets.

\subsection{Algorithm for OOD Sample Generation}

\noindent \textbf{Inputs and Outputs:} The algorithm requires a seed dataset $\mathcal{X}_{seed}$ and a sparsity parameter $k$ to define the extent of perturbation. The black-box model $f$ is used as a labeling oracle to record outputs, while the fingerprint data generation itself is independent of $f$. The expected outputs are two sets of pairs: $\mathcal{S}_{expl}$ containing perturbations following explainer-style distributions, and $\mathcal{S}_{ood}$ containing general out-of-distribution perturbations.

\noindent \textbf{Functional Overview:} This algorithm implements a Deterministic--Stochastic OOD Probing strategy to generate model fingerprints. It systematically explores a model $f$ by creating two perturbation sets from a seed set $\mathcal{X}$. The Explainer fingerprints ($\mathcal{S}_{expl}$) are generated by alternating between SHAP-style masking and LIME-style jittering. Conversely, the OOD fingerprints ($\mathcal{S}_{ood}$) utilize non-explainer perturbations, including group feature swapping with reference points and the addition of Gaussian noise $\mathcal{N}(0, \text{Var}(\mathcal{X}^{(i)}))$. This dual-pronged approach allows for the detection of discrepancies in model behavior when subjected to the specific sampling manifolds used by interpretability frameworks versus general out-of-distribution noise.

\subsection{Feature Rank Analysis:}
\vspace{-3mm}
\begin{figure}[!htbp]
\centerline{\includegraphics[width=1.03\columnwidth]{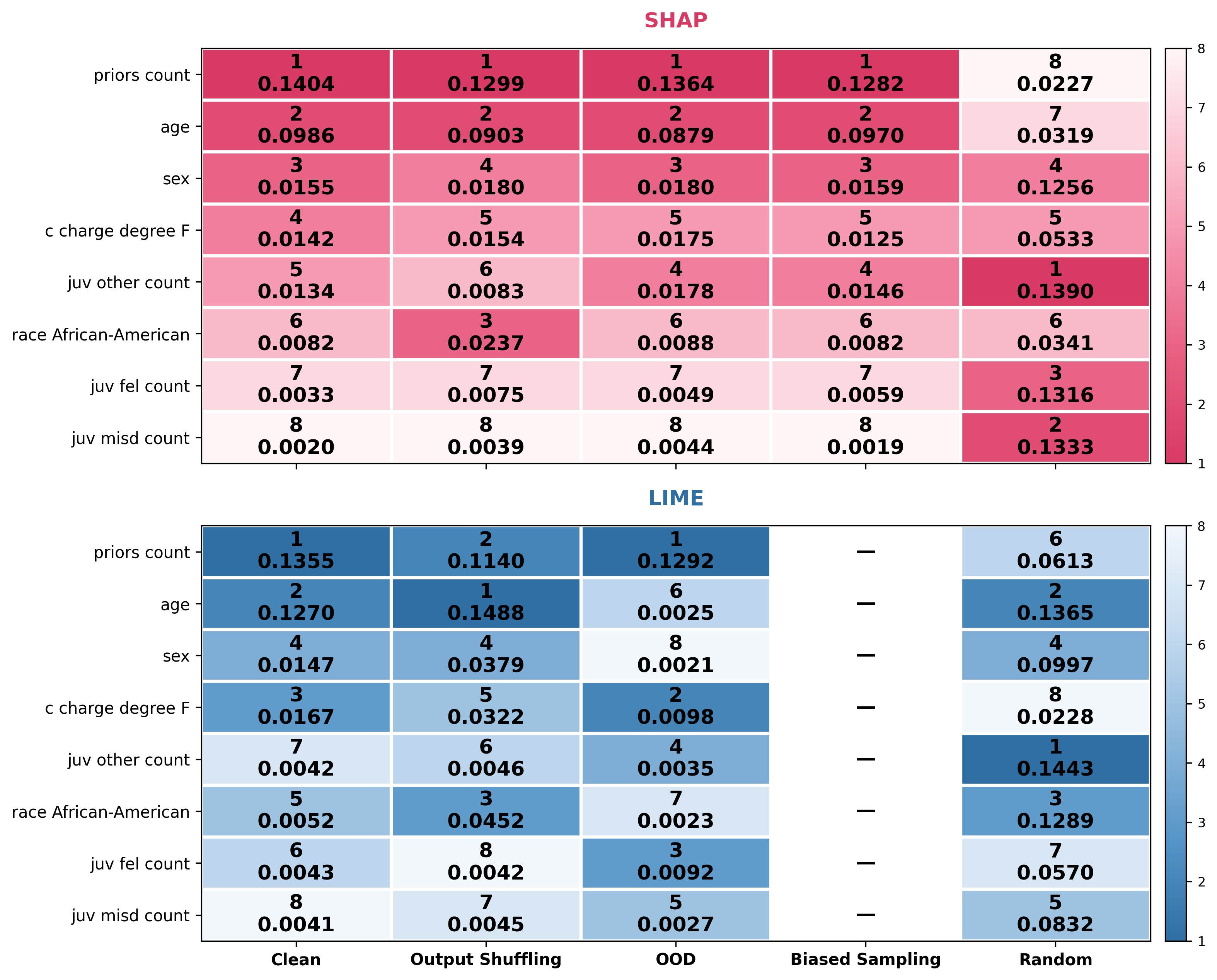}} 
\caption{Feature attribution heatmap along with rank indicator for all 5 experiment scenarios}
\label{fig:heatmap}
\end{figure}

As illustrated in Figure~\ref{fig:heatmap}, both LIME and SHAP exhibit consistent feature ranks under the clean baseline setup. However, under adversarial attack scenarios, LIME suffers severe fluctuations in both attribution values and feature rankings, confirming the successful execution of the attacks. The random control case reflects purely randomized feature attributions and rankings. Conversely, while SHAP demonstrates greater overall robustness than LIME, it appropriately captures a notable rank shift for the feature \texttt{sex} under the output shuffling attack. Note that the biased sampling heatmap for LIME is omitted ($N/A$), as the attack specifically targets the SHAP framework.
 
\end{document}